# Water induced bandgap engineering in nanoribbons of hexagonal boron nitride

*Chen Chen[†], Yang Hang[†], Hui Shan Wang, Yang Wang, Xiujun Wang, Chengxin Jiang, Yu Feng, Chenxi Liu, Eli Janzen, James H. Edgar, Zhipeng Wei, Wanlin Guo, Weida Hu\*, Zhuhua Zhang[\*], Haomin Wang[\*], and Xiaoming Xie*

C. Chen, H. S. Wang, X. Wang, Y. Feng, C. Liu, H. Wang
National Key Laboratory of Materials for Integrated Circuits, Shanghai Institute of Microsystem and Information Technology, Chinese Academy of Sciences, Shanghai, China;
Center of Materials Science and Optoelectronics Engineering, University of Chinese Academy of Sciences, Beijing, China;
CAS Center for Excellence in Superconducting Electronics (CENSE), Shanghai, China;
E-mail: hmwang@mail.sim.ac.cn

Y. Hang, W. Guo, Z. Zhang
Key Laboratory for Intelligent Nano Materials and Devices of Ministry of Education, State Key Laboratory of Mechanics and Control of Mechanical Structures, and Institute for Frontier Science, Nanjing University of Aeronautics and Astronautics, Nanjing 210016, China
E-mail: chuwazhang@nuaa.edu.cn

Y. Wang, W. Hu
State Key Laboratory of Infrared Physics, Shanghai Institute of Technical Physics, Chinese Academy of Sciences, Shanghai, China
E-mail: wdhu@mail.sitp.ac.cn

C. Jiang
National Key Laboratory of Materials for Integrated Circuits, Shanghai Institute of Microsystem and Information Technology, Chinese Academy of Sciences, Shanghai, China;
School of Physical Science and Technology, ShanghaiTech University, Shanghai, China

E. Janzen, J. H. Edgar
Tim Taylor Department of Chemical Engineering, Kansas State University, Manhattan, KS, 66506 USA

Z. Wei

State Key Laboratory of High Power Semiconductor Lasers, Changchun University of Science and Technology, Changchun, People's Republic of China

† These authors contributed equally to this work.

Different from hexagonal boron nitride (*h*BN) sheets, the bandgap of *h*BN nanoribbons (BNNRs) can be changed by spatial/electrostatic confinement. It has been predicted that a transverse electric field can narrow the bandgap and even cause an insulator-metal transition in BNNRs. However, experimentally introducing an overhigh electric field across the BNNR remains challenging. Here, we theoretically and experimentally demonstrate that water adsorption greatly reduces bandgap of zigzag oriented BNNRs (zBNNRs). *Ab initio* calculations show that water adsorbed beside the BNNR induces a transverse equivalent electric field of over 2 V/nm thereby reducing its bandgap. Field effect transistors were successfully fabricated from zBNNRs with different widths. The conductance of zBNNRs with adsorbates of water could be tuned over 3 orders in magnitude via electrical field modulation at room temperature. Furthermore, photocurrent response measurements were taken to determine the optical bandgap in zBNNR. Wider zBNNRs exhibit a bandgap down to 1.17 eV. This study yields fundamental insights in new routes toward realizing electronic/optoelectronic devices and circuits based on hexagonal boron nitride.

## 1. Introduction

Hexagonal boron nitride (*h*BN) is a wide bandgap semiconductor with a honeycomb lattice structure analogous to graphene.[1] It can act as an excellent insulating substrate and encapsulation layers for 2D electronic devices due to an atomically flat surface free of dangling-bonds and charge-impurities and superior chemical/thermal stabilities.[2,3] However, the large bandgap and insulating nature of *h*BN severely limit its potential applications in a wide range of electronic and optoelectronic devices.[4] Similar to other technologically-relevant semiconductors, the successful applications of *h*BN rely on the controlled fabrication of its nanostructures and the effective modulation of electronic properties. Typically, *h*BN has to combine with other materials to produce functional devices. If the band structure of *h*BN could be directly tuned, then homogeneous devices could be directly integrated on a single *h*BN substrate,[5] opening a new avenue to developing *h*BN-based integrated circuits of atomic thickness.

To modify its properties, *h*BN can be tailored into one-dimensional nanoribbons along selectable crystallographic directions. Because of its binary nature, *h*BN nanoribbons (BNNRs) exhibit enriched edge terminations, manifested as different edge structures, passivation and chemical decorations, eventually generating a rich variety of electronic and magnetic properties.[6-8] However, the large bandgap of BNNRs has limited their nanoelectronics applications.[9] Extensive theoretical efforts have been devoted to explore various strategies for modulating the bandgap of BNNRs, including physical and chemical methods. In particular, it has long been predicted that the bandgap of BNNRs can be significantly reduced by applying a transverse external electric field,[10,11] but no experimental verification has been reported.

The reason for this lies in twofold. First, the controlled fabrication of BNNRs with nanometer width is challenging. While the BNNRs can be realized by unzipping of *h*BN nanotubes,[12,13] their control of edge quality is poor. The fabrication of narrow BNNRs with selected crystallographic orientations and atomically smooth edges is also a daunting task. Second, to achieve an effective modulation of the bandgap of BNNRs, the transverse electric field needs to be as high as 1 V/nm.[10] Such an ultrahigh transverse electric fields are difficult to achieve and apply with traditional electrodes and, thus, hamper experimental realization.

Here, we report that the adsorption of water molecules can effectively modulate the bandgap of BNNRs. *Ab initio* calculations suggest a distinct gap modulation in zigzag-edged BNNRs (zBNNRs) when water molecules are favorably assembled along the edges to form an ice-like monolayer structure. Due to the concertedly accumulated polarity of water molecules, the ice-like water structure can induce a transverse electric field that effectively narrow the bandgap of zBNNRs. To confirm this result, zBNNRs were synthesized by catalytic cutting of *h*BN.[14] Electrical measurements obtained a high field effect on/off ratio of more than $10^3$ by introducing water vapor into zBNNRs prepared on a *h*BN substrate placed in a vacuum chamber. Optoelectrical measurements revealed that the bandgap of water-adsorbed zBNNRs is down to 1.17 eV from its intrinsic value of 4-5 eV. Our findings not only experimentally verify the giant Stark effect previously predicted for BNNRs but also provide a general approach for modulating electronic properties of low-dimensional semiconductors by making use of polar molecules towards device applications.

## 2. Results and Discussion

We first investigated the band structure of zBNNR in absence of water adsorption. The computational model is shown in Figure S1a. The model includes a zBNNR in a width *w* of 2.46 nm on an *h*BN substrate. The edges of zBNNR are passivated with H atoms. The band structure of zBNNR is shown in Figure S1c. As shown in Figure S1c, the bare zBNNR is an insulator with a bandgap of 4.17 eV, which is basically consistent with that reported in the literature.[10] On this basis, the adsorption of hexagonally arranged water molecules on the edges of zBNNR is taken into consideration. Here, the water molecule arrangement of this structure is optimized in energy and can stably existed at room temperature, according to simulation of molecule dynamics. The optimized computational model is shown in Figure S1b. The calculation results show that the bandgap of zBNNR was significantly reduced to 2.83 eV after the introduction of water molecule (Figure S1d).

Width dependence of the bandgap of zBNNR is also explored here. In order to increase the efficiency of calculation, the underlying *h*BN substrate is removed. Please be noted that it does not affect the overall tendency of bandgap evolution. The model adopted for the calculation (**Figure 1**a) corresponds to the band structure and density of states (DOS) shown given in Figure 1b. The calculations showed that the bandgap of zBNNR adsorbed with water decreases monotonically with its width (Figure 1c). In addition, it is found that the bandgap of the adsorbed water molecules is always significantly larger than zBNNR. As such, the electrical behavior of zBNNR adsorbed with water molecules is mainly determined by zBNNR.

The bandgap reduction originates from the enhanced transverse potential difference introduced by the polar water molecular adsorption. The zBNNR consists of a B atom terminated edge ($ZZ_B$) and a N atom terminated edge ($ZZ_N$). Figure 1d shows the charge densities of HOMO (highest occupied molecular orbital) and LUMO (lowest occupied molecular orbital) at **Γ** and **Y** point of a 2.46 nm width zBNNR adsorbed with water. It is found that the introduction of water aggravates the charge asymmetric distribution. Figure 1e extracts the variation of the planar average potential across the zBNNR, it shows that the potential of the $ZZ_N$ edge is significantly higher than $ZZ_B$. It is obvious that the water adsorption introduces a potential energy difference of more than 5 eV between the two edges.

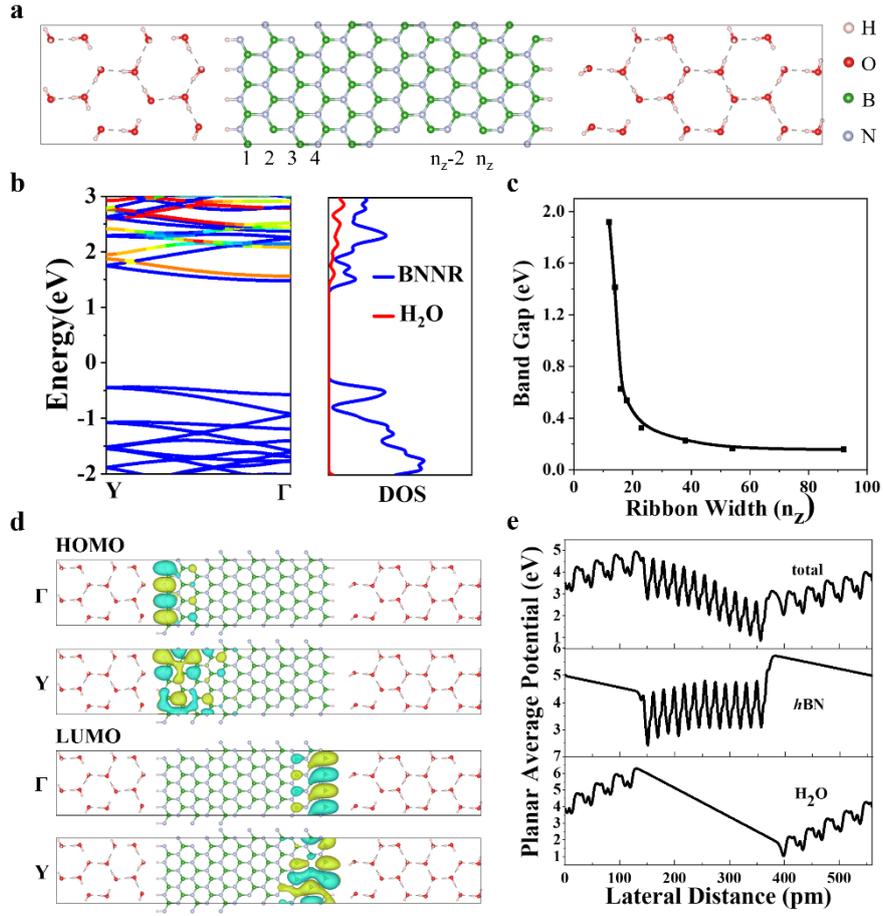

**Figure 1.** Band-gap modulation of BNNR by water adsorption. a) Diagram of the periodic arrangement of zBNNR and adsorbed water. The width of different zBNNRs is expressed in $n_z$. b) Band structure and density of states of the zBNNR with water adsorption. The color from blue to red represents the contributions from $h$BN to water. c) The bandgap of different width BNNRs under water adsorption. d) From up to down are the HOMO in Γ and Y points and LUMO in Γ and Y points. e) The planar average potential for zBNNR absorbed with water, water alone and zBNNR alone.

Due to the difference in electrostatic potential between B and N atoms, there is a spontaneous lateral electric polarization across zBNNR, and the polarization points from $ZZ_B$ to $ZZ_N$. it leads to that the potential of $ZZ_B$ is higher than that of $ZZ_N$. And therefore, the intrinsic zBNNR has a lower bandgap than $h$BN.[10] Nonetheless, the built-in electric field present in intrinsic zBNNR is still insufficient to reduce its bandgap to a sufficiently low level. Our calculation results show that the potential difference between $ZZ_B$ and $ZZ_N$ can be further reduced by introducing water adsorption at the edges, and then reduce the bandgap of zBNNR.

Adsorption kinetic was performed to investigate the evolution of water adsorption process at BNNR edges. Figure S2 shows the kinetic model of water molecules on the $h$BN surface. Figure S2a presents the evolution of lowest-energy atomic configurations when water molecules dock to zigzag edge of $h$BN. It is found that water molecules prefer to dock alternatively to a row, and then advance the next row. Figure S2b presents the corresponding free energy evolution during water adoption. It is found that the free energy barrier is less than 0.55 eV for self-assembly of water molecules in hexagonal arrangement along the $h$BN zigzag edge. After a few steps of molecule addition, the energy profiles evolve into a periodic up-down alternating level sequences. It indicates that a hexagonal phase of water molecules can exist stably along the zigzag edge of $h$BN even at room temperature.

Molecule dynamics simulation was carried out to investigate the stability of the hexagonal arrangement of water molecules on $h$BN at room temperature. The calculation results are shown in Figure S3. As shown in Figure S3, the whole system achieves equilibrium for about 7700 fs, indicating the stability of the adsorption. Hexagonal phase of two-dimensional (2D) ice is always experimentally observed on clean surfaces.[15-18]

Further theoretical calculation shows that the water molecules prefer to adsorb on the edge of zBNNR rather than the bulk surface. A model is set-up to describe the adsorption of water molecules on the surface and edge of $h$BN. The model includes a water molecule, a $h$BN bulk surface and a ZZ oriented $h$BN edge upon $h$BN surfaces. The results are given in Figure S4. As shown in Figure S4, the total energy of water adsorption on the $h$BN bulk surface is -2646.9246 eV, while the total energy of adsorption on the $h$BN edge is -2647.0296 eV. This indicates that water molecules are more likely to adsorb at the edges of zBBNR.

The zBNNRs are fabricated via crystallographically cutting of catalytic nano-particles. Zn nano-particles can cut $h$BN and produce nano-trenches along zigzag direction at $H_2$ atmosphere and high temperature.[14] The typical width of the trenches is about 5 nm. As shown in **Figure 2**a, zBNNRs can be found between two parallel trenches. The zBNNRs fabricated by this approach possess sharp edges and are along a specific crystallographic orientation. The width of zBNNR can be examined by AFM. Figure 2c is the AFM height image of zBNNR with different width, which width are 6, 23, 52, 128 nm. More AFM images of zBNNRs with different width are shown in Figure S5. This approach proved to be an efficient method to fabricate high-quality zBNNRs with tunable widths.

Field effect transistor devices were fabricated to explore the electrical properties of zBNNRs, as shown in Figure 2b and Figure S6. zBNNRs with different width were fabricated into back-gated FET devices with channel length of 1 μm. The process of "metal transfer and etching" in device fabrication ensures a good electrical contact between zBNNRs and metal electrodes, which is particularly important to maintain the performance of the device.[19] The detailed device fabrication processes can be described in method section.

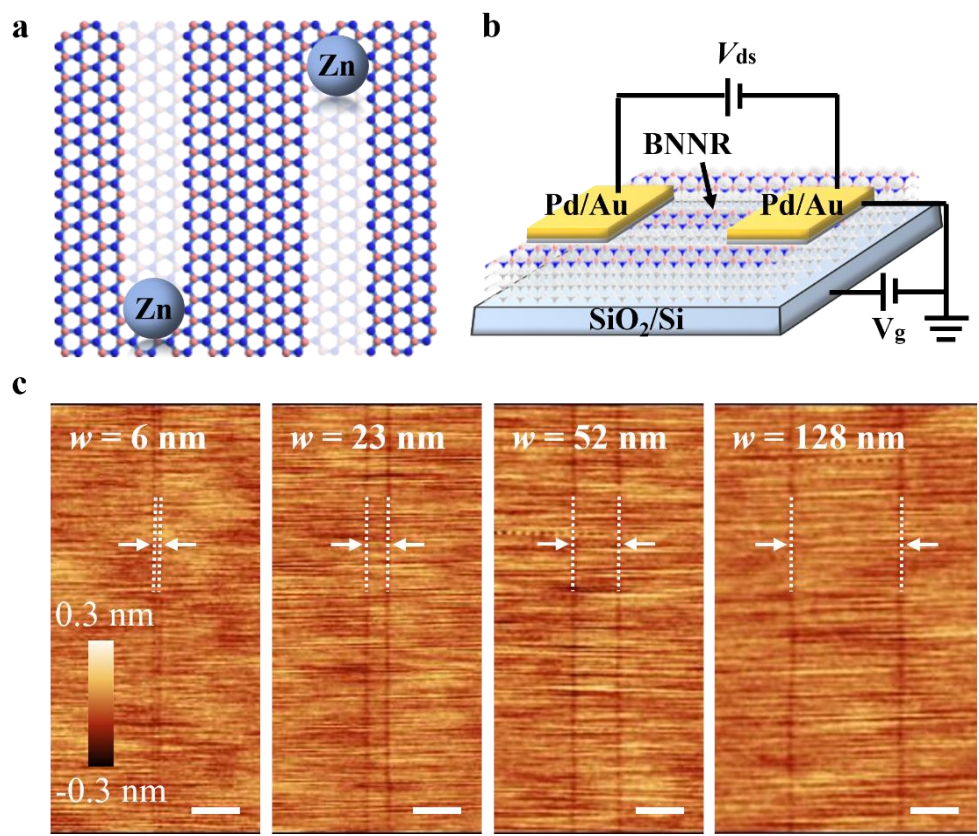

**Figure 2.** zBNNR fabrication through nanoparticles catalytic etching. a) A schematic illustrated the zBNNR between two parallel trenches etched by Zn nanoparticles on the surface of *h*BN. b) A schematic diagram of the zBNNR device for electric measurements. c) Atomic force microscopy (AFM) height images of BNNRs with different width. The scalebar is 50 nm.

Since zBNNRs were divided by two *h*BN trenches, it is possible for water molecules to fill in the *h*BN trenches spontaneously, forming zBNNRs with two edges are both adsorbed by water. To confirm that BNNRs can indeed be adsorbed by water in the environment, the as prepared zBNNR devices were placed in a vacuum (~$10^{-4}$ Pa) probe station and baked for 2 h to remove the adsorption of water at the zBNNRs edges. As shown in **Figure 3**a, the zBNNRs without any adsorption has no response to the back-gate voltage, and the source drain current $I_{ds}$ are only in the order of 1 pA. Then, after passing in argon with water vapor and keeping for 12 h, it can be found that zBNNR exhibits obvious switching modulation to gate voltage, and the on-state current is also significantly increased more than two orders in magnitude. At the same time, we found that the water vapor in the air environment can also produce effective adsorption to the zBNNR edges. Therefore, in this work, zBNNR devices are both tested in air environments.

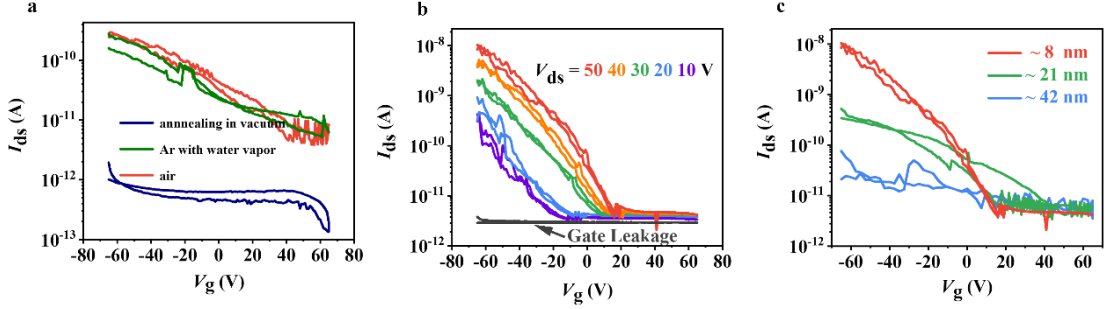

**Figure 3.** Electronic transport through zBNNR at room temperature. a) Electric transport of the zBNNR device under different atmosphere. The red line represents the transport curve measured in air while the navy-blue line represents the transport curve measured in vacuum (~$10^{-4}$ Pa) after baking. The green line represents the electric transport of the zBNNR devices after the vacuum chamber was filled in Ar gas with water vapor. b) The transfer curve under different source-drain voltage ($V_{ds}$). A typical curve of gate leakage current is also plotted for comparison. c) The transfer curves of zBNNR with different width.

Although the direct observation of water adsorption on the *h*BN surface is indeed difficult, we can still use scanning probe and synchrotron radiation technology to further characterize the properties of water on the surface of *h*BN. We performed x-ray absorption spectrum (XAS) to probe the water adsorption on the *h*BN surface. As shown in Figure S7, XAS has two significantly absorption peaks in 536 eV and 542 eV, which correspond to the disordered and crystallized water, respectively.[20,21] The adsorption water mainly existed in a crystallized form. In our experiments, it is found that water preferentially fills in the edges and trenches. We monitored the evolution of surface morphology of *h*BN over times, and we found water adsorption on the surface of *h*BN. When the sample with BNNRs was placed under air environments, the coverage of water on the *h*BN surface gradually increases over time and uniformly covers the *h*BN surface, as shown in Figure S8.

The conductance of zBNNR can be modulated by gate voltage. Typically, $h$BN is an insulator with a bandgap of about 6 eV. However, water adsorbed zBNNR can exhibit a semiconductor behavior. Figure 3b shows a transfer curve of an 8 nm width zBNNR device measured at 300 K under $V_{ds}$ from 10 V to 50 V and back-gate voltage $V_g$ from -65 V to 65 V. The device exhibits excellent switching characteristics. A high on/off ratio was measured more than $10^3$. This is the first experimental report on gate voltage modulation of zBNNR transistor. The device was turn on under negative gate voltage and turn off under positive gate voltage, which means that zBNNR transistors are electron-dominated n-type semiconductors. As the $V_{ds}$ increases, the on/off ratio of the zBNNR transistor also increases. For comparison, the gate leakage current measured is ~2 pA, which is comparable to the off current. The absence of hysteresis in transfer curve implies that the charge trapping rarely happens in zBNNR devices. The relationship between drain-source current $I_{ds}$ verses $V_{ds}$ was also measured, and the results are given in Figure S9. The measured on-state current through zBNNR are 10 nA at $V_{ds}$ = 50 V. The presence of a Schottky barrier at the Pd-BNNR interface is evident by the non-linear behavior of the *I-V* curve. The current density of zBNNR can reach 1 μA/μm, which may benefit from less scattering of carriers at zBNNR edges. The breakdown measurements were performed in a BNNR FET device with 10 nm channel width. As shown in Figure S10, the breakdown results shows that our BNNR devices can stand very high $V_{ds}$ up to 86 V. The breakdown current reaches about 800 nA. This implies the potential of BNNR in the application of high-power devices.

The electric properties of zBNNR also depend on their width. As shown in Figure 3c, zBNNRs wider than 42 nm rarely exhibit semiconductor switching characteristics. This is consistent with the insulating nature of $h$BN sheet. Similarly, we did not observe any similar semiconducting result in naturally formed $h$BN step edge and $h$BN trenches. This means that the semiconducting properties of the devices are derived from zBNNR.

Wider zBNNR devices exhibit significant photocurrent response to light illumination, despite their relatively small gate on-off ratios, as shown in **Figure 4**a. This is consistent with our calculated results. We extracted the response of zBNNR with different widths to gate voltage, and found that zBNNR devices with a width greater than 20 nm are basically in the off state. However, as the width increases, the photocurrent on/off ratio of zBNNR increases significantly. For zBNNR with a width of ~42 nm, the photocurrent light to dark ratio can reach about 150 under the irradiation of 1060 nm near-infrared light (region II). Figure 4b shows the time-dependent photocurrent in two zBNNR devices under illumination with a 1060 nm laser in power of 35 mW. Their widths are 33 nm and 8.5 nm, respectively. The dark *I-V* curves of the 33 nm width devices were shown in Figure S11. The $I_{ds}$ are not sensitive to the changes of $V_g$. With further increase in width, the properties of zBNNRs exhibit bulk-liked properties, that is, good electrical insulation and insensitivity to light illumination (region III).

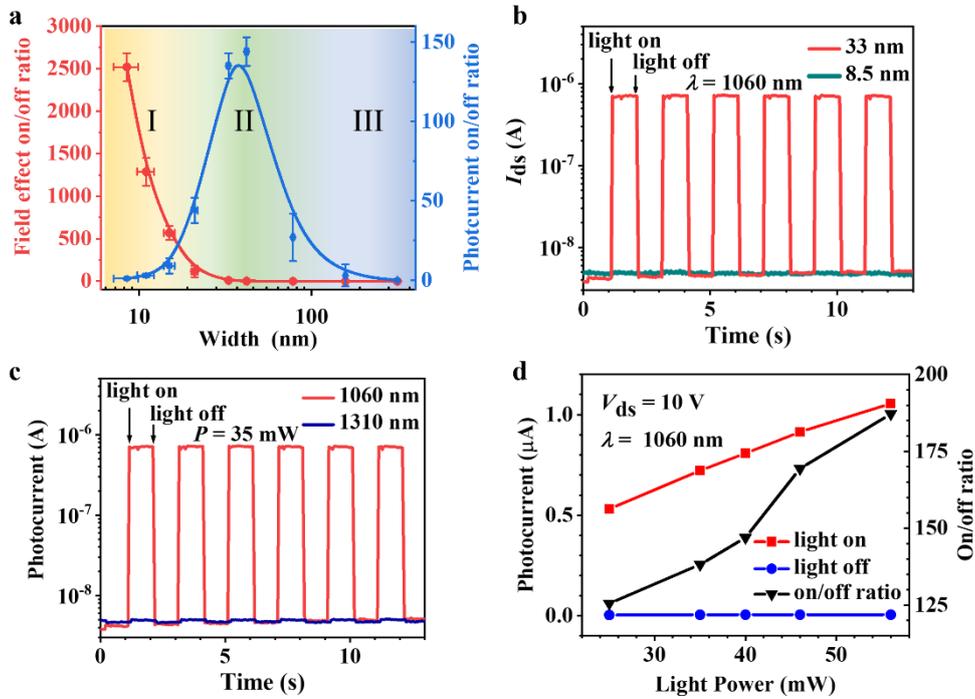

**Figure 4.** Optoelectronic characteristics of zBNNR devices. a) Field effect and current on/off ratio of the devices versus the width of zBNNR. b) Time-dependent photocurrent in two zBNNR devices under illumination with a 1060 nm laser in power of 35 mW. Their widths are 33 nm and 8.5 nm, respectively. c) Time-dependent photocurrent of the device under illumination with a 1060 nm and 1310 laser in a power of 35 mW. d) Photocurrent of a zBNNR device with a width of 33 nm under different power laser irradiation with a wavelength of 1060 nm.

To determine the minimum bandgap of zBNNRs, we measured the photo-response currents of a zBNNR device (the width is 33 nm) under laser illumination of different wavelengths. As shown in Figure 4c, the light absorption wavelength of zBNNR device is between 1060 nm and 1310 nm. This corresponds to an optical bandgap between 1.17 eV and 0.95 eV. Photocurrent of the device under light illumination in different wavelength (550 to 1050 nm) was also measured (Figure S12). The photocurrent on/off ratio of the device increased linearly with the increase of light power density, and the photoresponsivity was 16.9 nA/mW, as shown in Figure 4d.

The water layer adsorbed BNNR is quite stable under an ambient condition. We measured the time-stability of the samples under a baking temperature of 50°C. As shown in Figure S13a, we found that BNNR devices kept a high on/off ratio over 10 days in high vacuum environment. We also measured the current on/off ratio after baking at various temperature for 30 min in a vacuum environment ($10^{-4}$ Pa). As shown in Figure S13b, a high on/off ratio can be well maintained after 100°C baking. The adsorbed water can only be removed after high temperature baking (over 200 °C for 2 hours). It indicates that the stability of BNNR devices can be further improved by suitable packaging and encapsulation.

The bandgap modification of zBNNR originates from the asymmetric rearrangement of adsorbed water, which is induced by the asymmetic edge structure of zBNNR. Due to the structural asymmetry of the two edges of zBNNR, the arrangement of water molecular on these two edges is also different. This difference further magnifies the difference in potential energy at the two edges. For armchair oriented BNNR (aBNNR), its two edges are completely symmetrical and cannot induce asymmetric arrangement of adsorbed water. Therefore, we cannot observe the band structure modulation induced by water adsorption in aBNNR devices. We also prepared AC oriented BNNR devices, and the electrical test results were consistent with our predictions, as shown in Figure S14. They always show insulting behavior.To the zigzag oriented BNNR, the break of the lattice symmetry induces a spontaneous transverse electric polarization due to the existence of the asymmetrical ionic potential, which makes zBNNRs obviously different from armchair oriented BNNR. For BNNRs with other special chirality, similar phenomena can be expected if their two edges differ in symmetry. However, due to the difficulty in the preparation of special chiral BNNRs, the experimental verification is difficult.

We also noted that, the theoretically calculated bandgap of zBNNR adsorbed with water is significantly smaller than the experimental results. It is mainly due to the fact that DFT calculations always underestimate the bandgap of semiconductors and insulators.[22] Nevertheless, the calculated results can still accurately reflect the trends of water adsorption on the bandgap variation of zBNNR. Furthermore, compared with the modeling structure for calculations, the underlying $h$BN substrate of measured zBNNR devices is thicker (>10 nm). Generally, the dielectric constant of thicker $h$BN is smaller, which also caused the differences to calculated results.[23] The increase in the thickness of underlying $h$BN substrate in the computational model always cause the increase of the bandgap of zBNNR (see Figure S15).

## 3. Conclusion

A systematic study was carried out to investigate the bandgap modulation of water adsorbed zBNNR. Our calculations show that the water adsorption at zBNNR edges can induced strong transverse electric field, thus narrowed the bandgap of zBNNR. Through the adsorption of water at the edges, we measured the gate modulated transport of zBNNR devices for the first time. The bandgap is tunable by the width of zBNNR, which is benefit to the homo-integration of 2D electric and opto-electric devices based on *h*BN. This provides a new route toward practical electronic devices based on boron nitride materials.

## 4. Methods

**zBNNR fabrication.** Fresh *h*BN flakes were cleaved onto quartz substrates mechanically, followed by a 650°C annealing in an $O_2$ flow for 60 min to remove organic residues on the surface. $ZnCl_2$ solution (0.05 g/L) was first spun at 4,000 r.p.m. for 100 s onto the surface of *h*BN flakes. The substrates were then baked for 10 min. The etching processes were produced at a vacuum chamber under $Ar/H_2$ (50:25 sccm) atmosphere at high temperature (1300°C) for 60 min. After etching, the chamber was naturally cooled to room temperature in the protection of argon. zBNNRs are formed between two parallel trenches on the surface of *h*BN, and can be found by SEM and AFM imaging

**Device fabrication of zBNNR FET devices.** The *h*BN flakes with zBNNRs were transferred onto Si substrate with 300 nm $SiO_2$ by metal assisted transfer method.[24] As shown in Figure S16, a 50 nm thickness Pd layer is firstly deposited onto *h*BN/quartz substrate surface by e-beam evaporation. A layer of PMMA (950 A2) is spin-coated on the top of the Pd film (3000 rpm, 60 s) and baked at 150 °C for 5 min as a sacrificial layer to prevent tape residue contamination. The PMMA/Pd/*h*BN layer is picked up by Nitto tape as handle layer, and then pressed onto a silicon substrate with 90 nm $SiO_2$ layer. The substrate was then heated in 60°C for 10 min. And then, the Nitto tape and PMMA layer are removed by socking in acetone, leaving Pd covered *h*BN with zBNNRs on $Si/SiO_2$ surface. The source and drain electrodes were patterned by standard e-beam lithography method. A layer of Au with 50 nm thickness was then deposited on the electrode patterns by e-beam evaporation. After removing the organic resist, the sample was soaked into a Pd etchant (Transene Palladium etchant TFP) for 20 s. Finally, the electrodes of Au/Pd are formed on zBNNR.

**Electrical and opto-electrical measurements of zBNNR FET devices.** The electrical transport was measured via a Keithley 4200 semiconductor parameter analyzer in an atmosphere-tunable probe station at room temperature. The photocurrent measurements were performed in ambient conditions using confocal microscopy and an Agilent 2902 semiconductor parameter analyzer.

**AFM measurements.** Park atom force microscope (AFM) was used to perform the AFM measurements. The contact probes have a nominal radius of ~25 nm (CONTSCR for tapping mode scanning and CONTSCR for contact mode scanning).

**Theoretical calculations.** First-principles calculations were performed by using Vienna Ab initio Simulation Package with the Perdew-Burke-Ernzerhof-type generalized gradient approximation for the exchange-correlation functional and the projector augmented wave method. All the atoms are relaxed until the residual force was less than 0.02 eV/Å. A $1 \times 3 \times 1$ grid of k points and a plane-wave cutoff energy of 400 eV are used for the self-consistent calculations and electronic structure analysis. To eliminate interactions between the neighboring cells of slab models, the vacuum region is fixed to 15 Å. In all calculations, the $D_2$ method is adopted to describe the van der Waals interactions.

**X-ray absorption spectroscopy**

The X-ray absorption spectroscopy (XAS) experiments were performed at beamline 02B at the Shanghai synchrotron radiation facility (SSRF) in China. 02B is a bending magnet beamline equipped with monochromator composed of one water-cooled plane mirror and three varied-line-spacing gratings, which can provide ~1011 photons per second in the energy range from 40 to 2000 eV with the average energy resolution of 4000. O K-edge of the samples were measured in total electron yield (TEY) mode under ultrahigh vacuum ($10^{-9}$ Torr) with an incident X-ray beam size of $0.1 \times 0.2$ mm$^2$. The spectra were normalized to the photon flux of incident beam monitored by the Au mesh. The photon energy was calibrated with the spectra of $SrTiO_3$.

**Supporting Information**

Supporting Information is available from the Wiley Online Library or from the author.

**Acknowledgements**

C.C. and Y.H. contributed equally to this work. H.W. and X.X. directed and supervised the research work. H.W. conceived and designed the research work. C.C. and H.S.W. performed the etching processes on *h*BN. C.C. performed the AFM measurements of BNNRs. Y. H. and Z. Z. performed the modeling and DFT calculations. Z.W., P.W. and W.H. carried out the opto-electric measuements. C.C., H.S.W., C.J., X.W., Y.F., C.L. analyzed the experimental data. H.W., Z. Z. and W. H. contributed to critical discussions of the manuscript. J. H. E. fabricated the *h*BN crystals. C.C. and Y. H. co-wrote the manuscript. All the authors contributed to critical discussions of the results and manuscript.

This work was supported by the Strategic Priority Research Program of Chinese Academy of Sciences (XDB30000000), National Natural Science Foundation of China (91964102, 51772317, 12004406), the National Key R&D Program of China (2022YFF0609800, 2017YFF0206106), the Science and Technology Commission of Shanghai Municipality (20DZ2203600), China Postdoctoral Science Foundation (BX2021331, 2021M703338), Shanghai Post-doctoral Excellence Program(2021515) and Soft Matter Nanofab (SMN180827) of Shanghai Tech University, Natural Science Foundation of Jiangsu Province (BK20220872), Jiangsu Funding Program for Excellent postdoctoral talent (2022ZB232). Support for *h*BN crystal growth was provided by the USA Office of Naval Research, award number N000142012474.

Received: ((will be filled in by the editorial staff))
Revised: ((will be filled in by the editorial staff))
Published online: ((will be filled in by the editorial staff))

Supporting Information

**Water induced bandgap engineering in nanoribbons of hexagonal boron nitride**

*Chen Chen†, Yang Hang†, Hui Shan Wang, Yang Wang, Xiujun Wang, Chengxin Jiang, Yu Feng, Chenxi Liu, Eli Janzen, James H. Edgar, Zhipeng Wei, Wanlin Guo, Weida Hu\*, Zhuhua Zhang\*, Haomin Wang\*, and Xiaoming Xie*

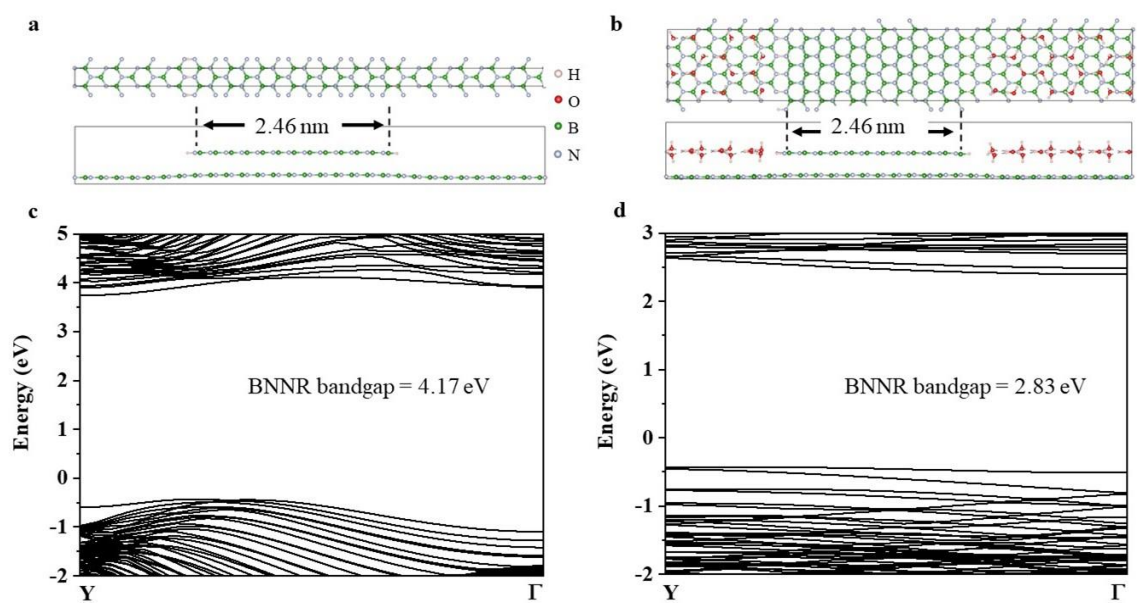

**Figure S1.** Band structures of zBNNRs under different edge adsorption of water. Diagram of a 2.46 nm width zBNNR a) without and b) with water adsorption at the edges. The longitudinal direction is the infinitely extended direction. The band structure of zBNNR in (a) and (b) is drawn in (c) and (d), respectively.

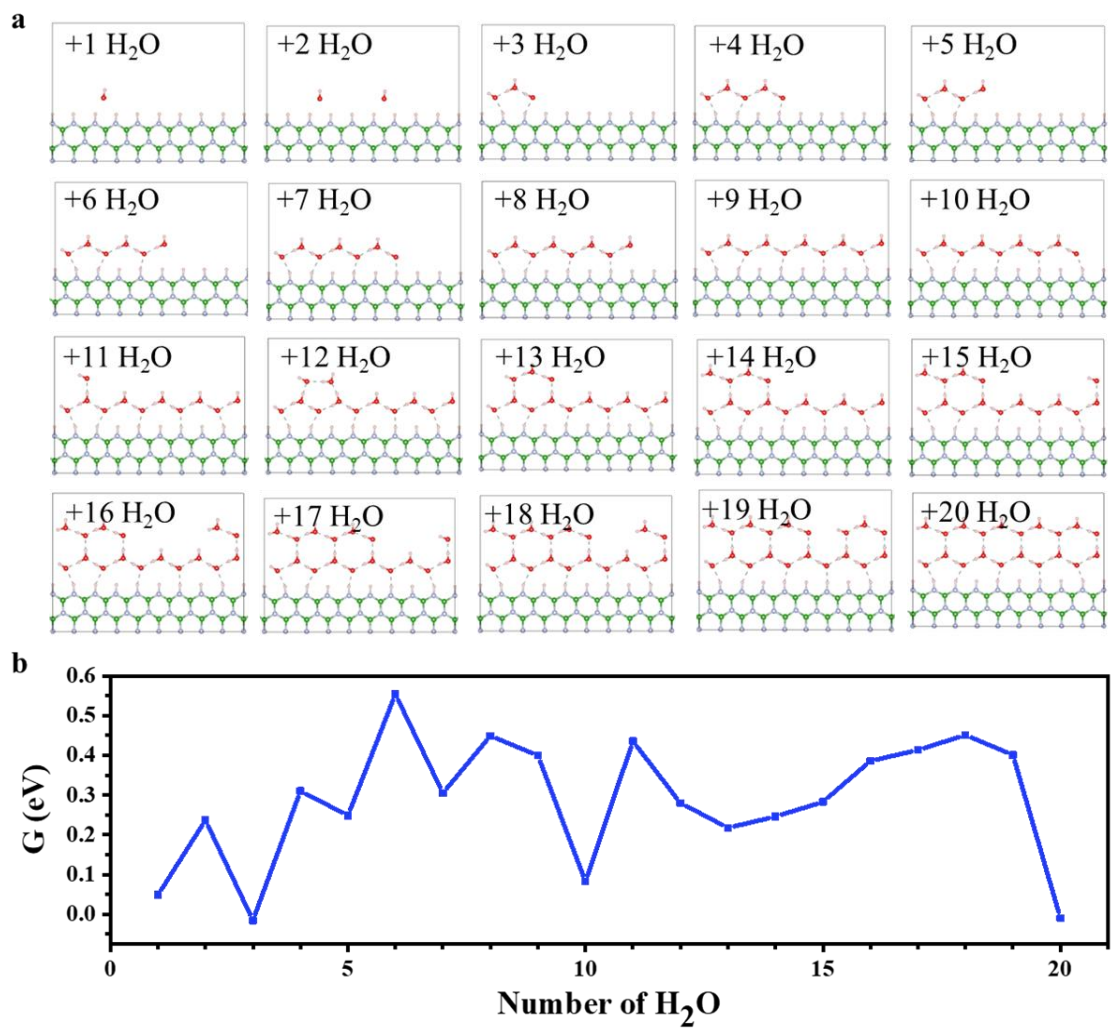

**Figure S2.** Kinetic models of water molecules with different configurations on the *h*BN surface. a) Optimal atomic configurations during the docking of water molecules to the zigzag *h*BN edges. b) Corresponding free energy evolution during water adoption.

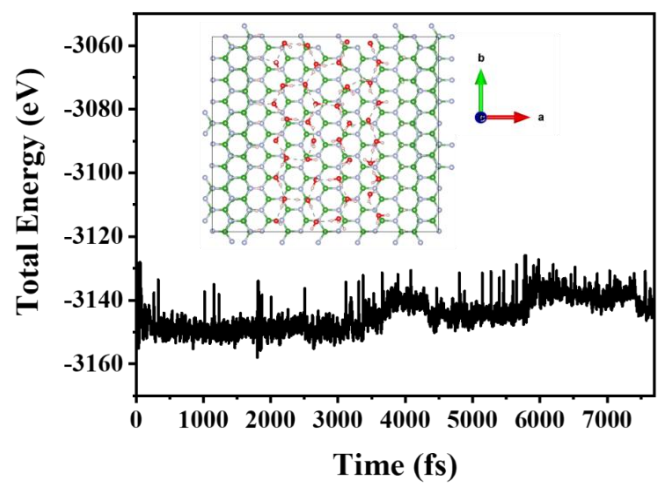

**Figure S3.** Molecule dynamics simulation of the hexagonal arrangement of water molecules on *h*BN at room temperature.

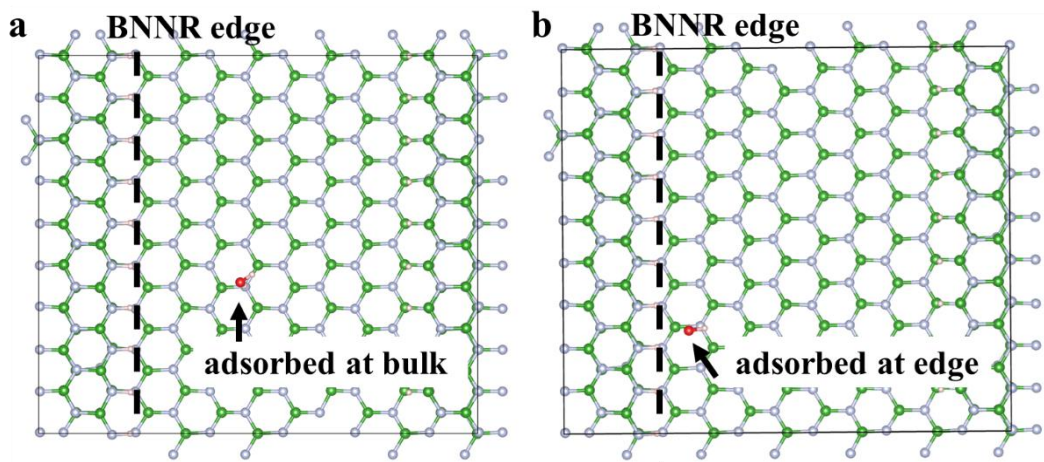

**Figure S4.** The total energy of *h*BN with water adsorbed at a) bulk and b) edge. The total energy is -2646.9246 eV and -2647.0296 eV, respectively.

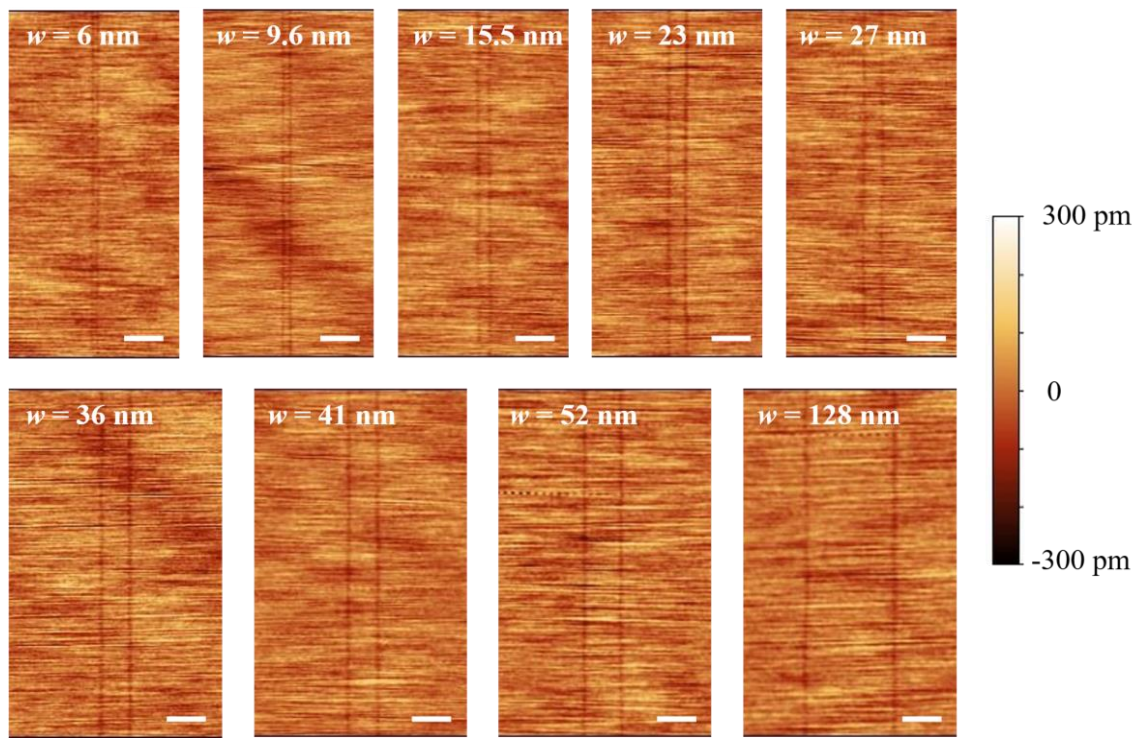

**Figure S5.** Atomic force microscopy height images of zBNNRs with different width. The scalebar is 50 nm.

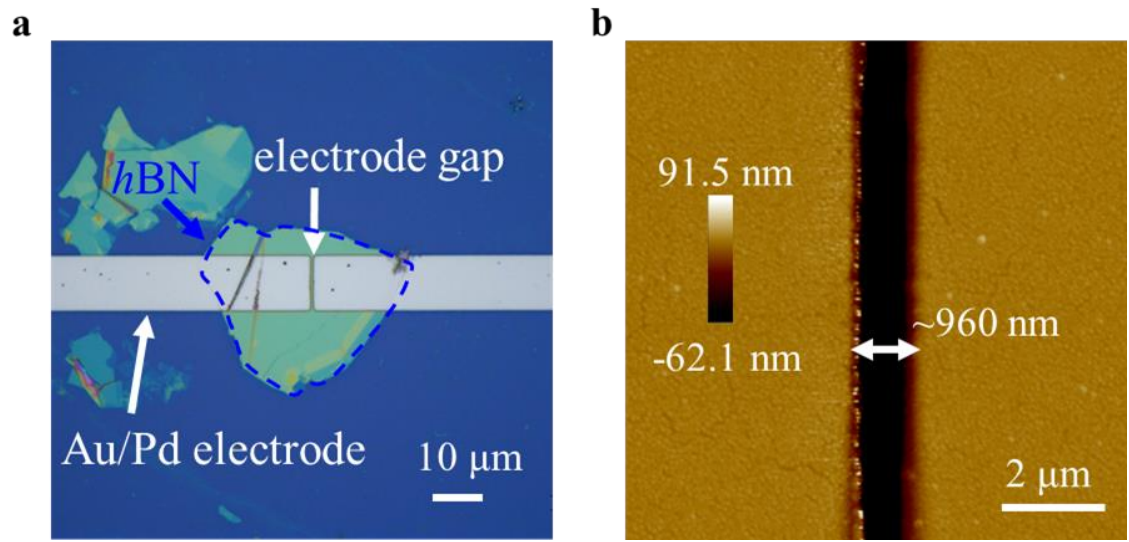

**Figure S6.** a) Optical image and b) zoom-in AFM height image of a zBNNR device.

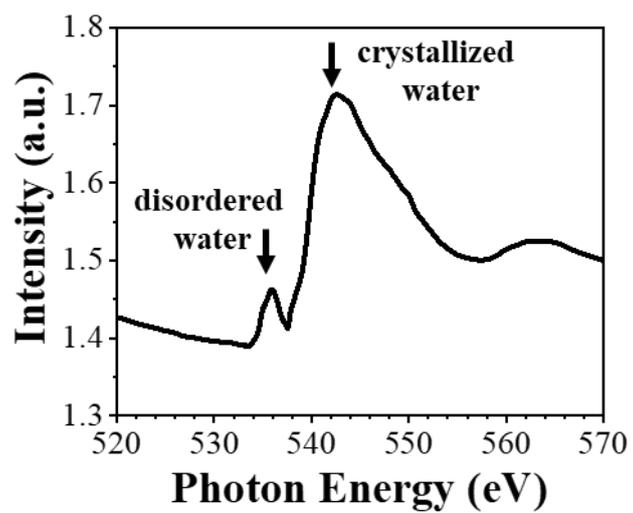

**Figure S7.** X-ray absorption spectrum of water adsorbed BNNR sample.

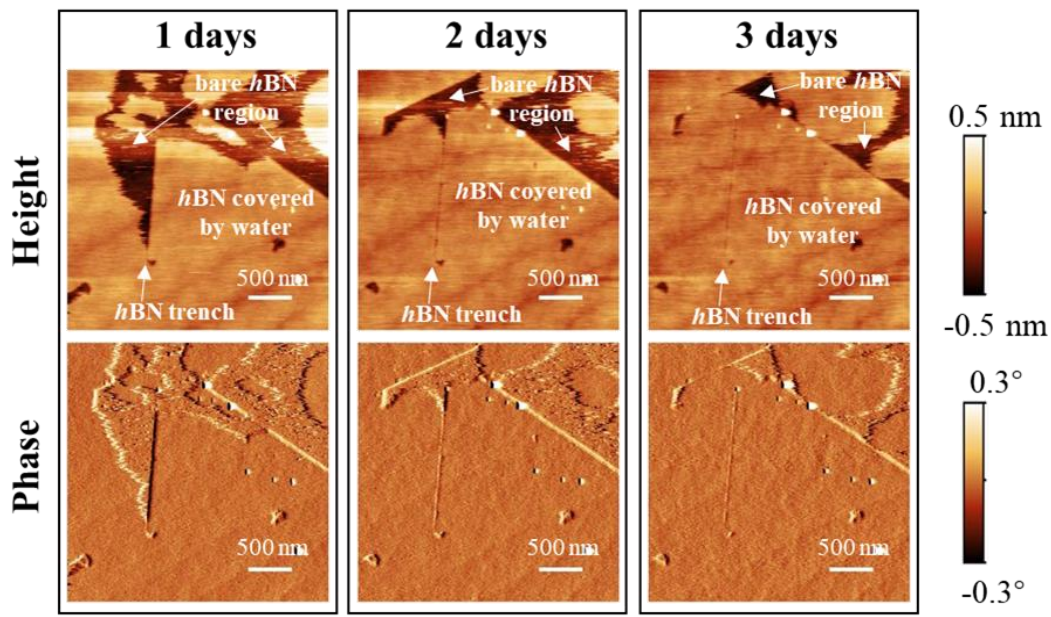

**Figure S8.** The surface morphology evolution of a BNNR sample over times.

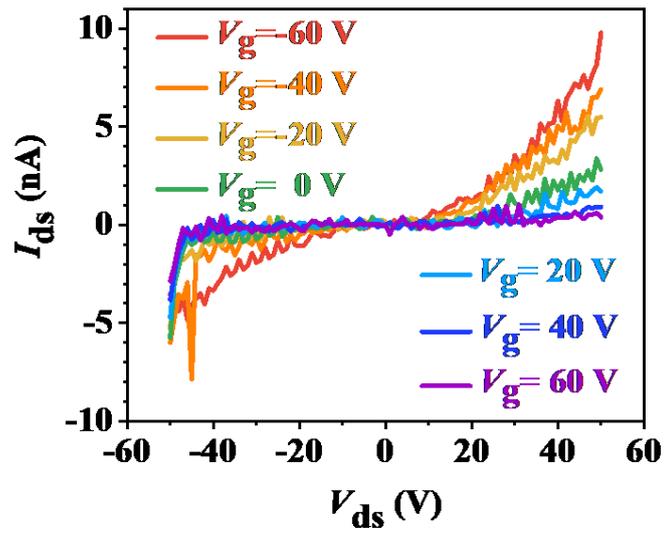

**Figure S9.** The output curve of an 8 nm width zBNNR device at room temperature.

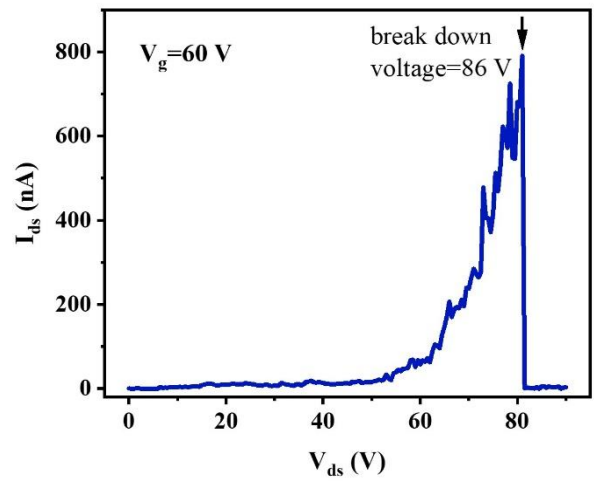

**Figure S10.** The electric breakdown curve of a FET device with 10 nm width BNNR as channel. The device was measured at on state with a *V*g is 60 V.

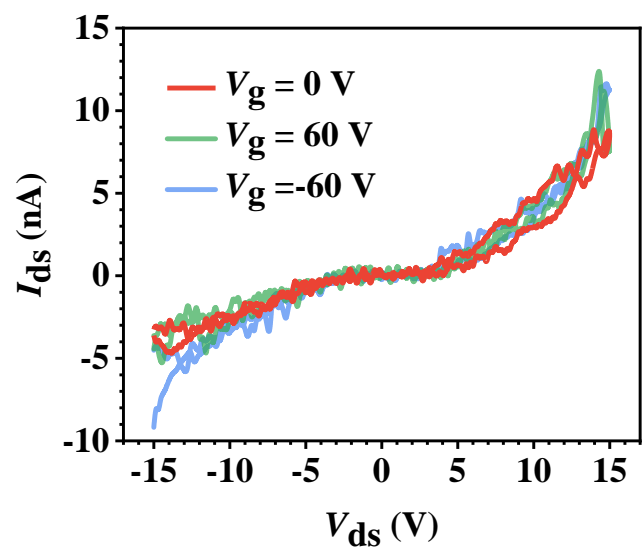

**Figure S11** The dark *I-V* curves of a 33 nm width zBNNR device under different gate voltages.

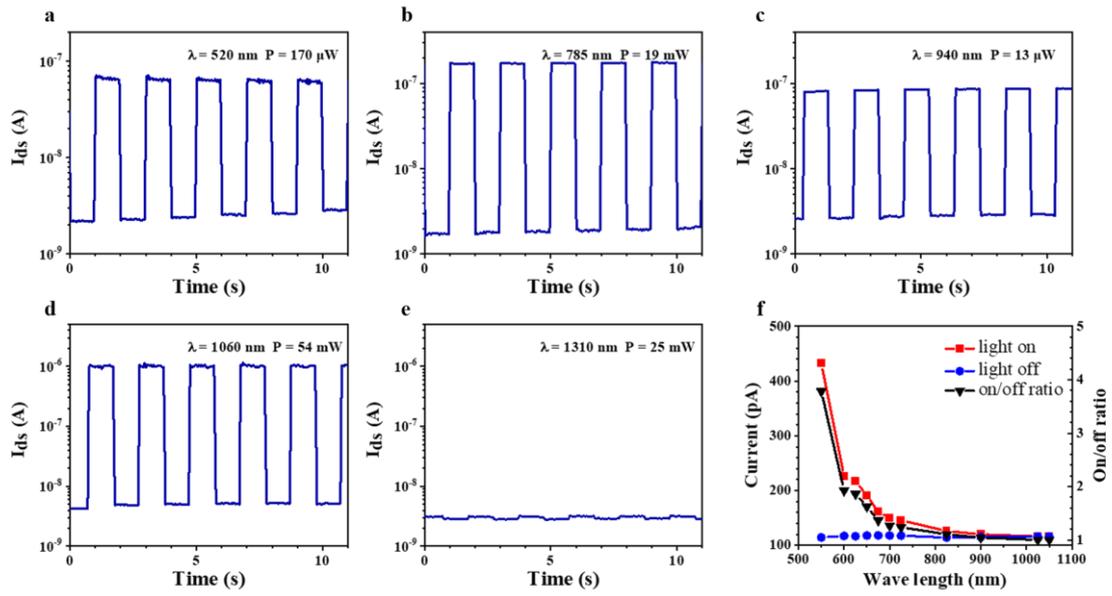

**Figure S12.** Time-dependent photocurrent in two zBNNR devices under illumination with different wavelength laser of a) 520 nm, b) 785 nm, c) 940 nm, d) 1060 nm and e) 1310 nm. f) Photocurrent of the device under laser illumination in different wavelength with a fixed power of 20 μW.

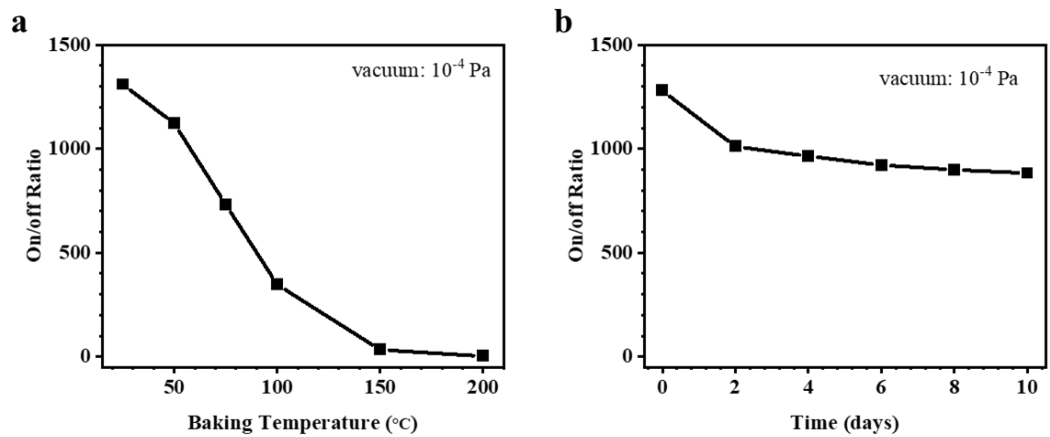

**Figure S13.** a) Time stability of a BNNR device with 10 nm channel width under the environment of $10^{-4}$ Pa environment. b) The on/off ratio after various temperature baking on vacuum with pressure less than $10^{-4}$Pa.

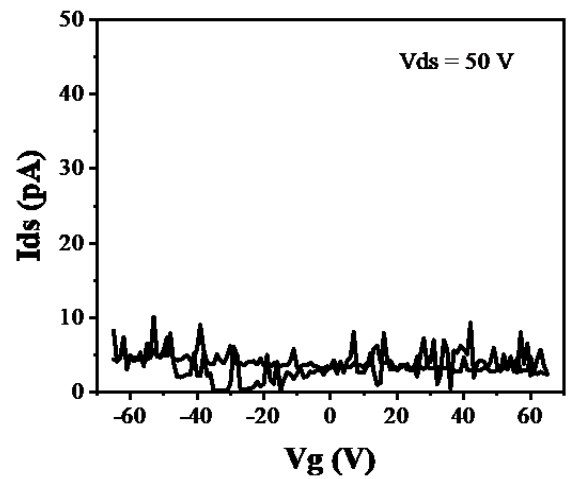

**Figure S14.** The transport curve of a typical armchair oriented BNNR device.

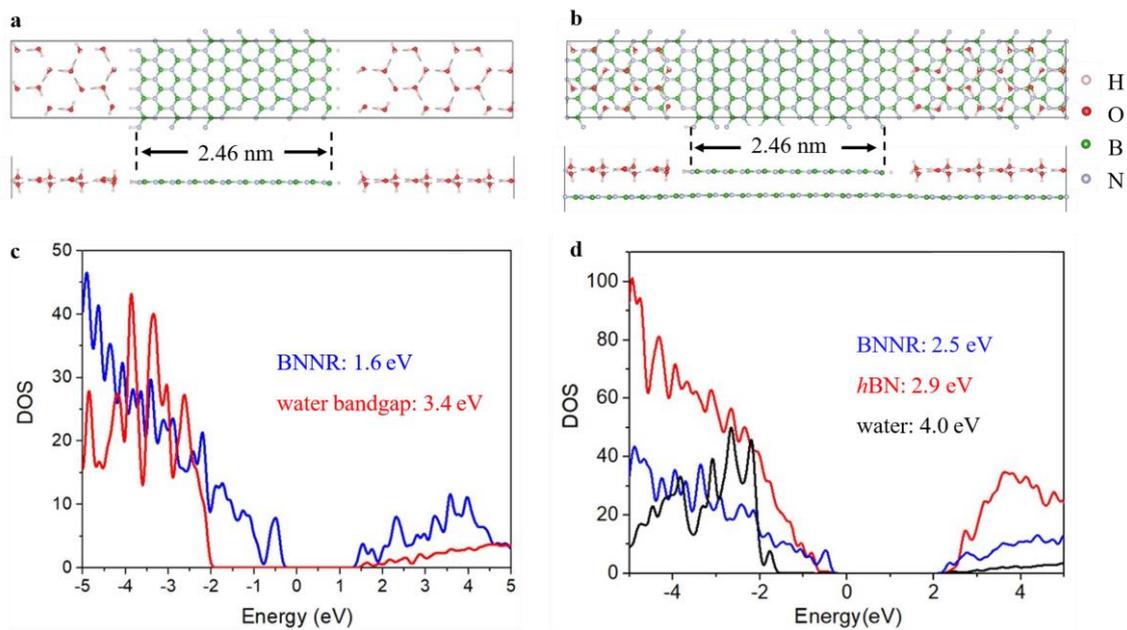

**Figure S15.** Band structures of water-adsorbed zBNNRs with different *h*BN substrate thicknesses. Structure schematics of a) single-layered and b) double-layered zBNNRs with a width of 2.46 nm. c) and d) are the band structures of zBNNRs in (a) and (b), respectively.

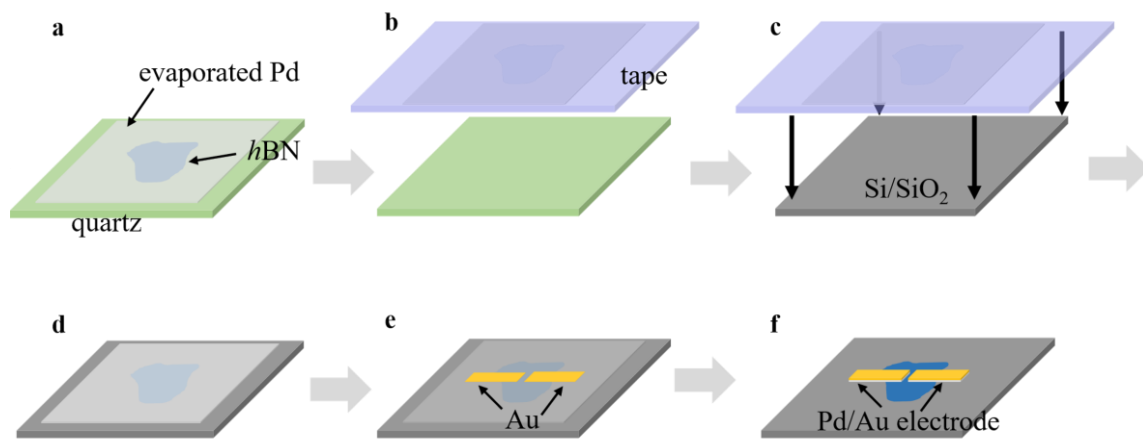

**Figure S16.** Schematic for zBNNR transferring and FET device fabricating.